\documentclass[twocolumn,aps,showpacs,prc,superscriptaddress,nofootinbib,
amsmath,floatfix]{revtex4}
\usepackage{graphicx}
\usepackage{bm}

\begin{document}

\title{Breaking of nucleon Cooper pairs at finite temperature in 
$^{\bm{93-98}}$Mo}

\author{K. Kaneko}
\affiliation{Department of Physics, Kyushu Sangyo University, Fukuoka  
813-8503, Japan}
\author{M. Hasegawa}
\affiliation{Laboratory of Physics, Fukuoka Dental College, Fukuoka 814-0193, 
Japan}
\author{U. Agvaanluvsan}
\affiliation{Lawrence Livermore National Laboratory, L-414, 7000 East Avenue, 
Livermore, CA 94551, USA}
\affiliation{North Carolina State University, Raleigh, NC 27695, USA}
\author{E. Algin}
\affiliation{Lawrence Livermore National Laboratory, L-414, 7000 East Avenue, 
Livermore, CA 94551, USA}
\affiliation{North Carolina State University, Raleigh, NC 27695, USA}
\affiliation{Triangle Universities Nuclear Laboratory, Durham, NC 27708, USA}
\affiliation{Department of Physics, Osmangazi University, Meselik, Eskisehir, 
26480 Turkey}
\author{R. Chankova}
\affiliation{Department of Physics, University of Oslo, N-0316 Oslo, Norway}
\author{M. Guttormsen}
\affiliation{Department of Physics, University of Oslo, N-0316 Oslo, Norway}
\author{A.C. Larsen}
\affiliation{Department of Physics, University of Oslo, N-0316 Oslo, Norway}
\author{G.E. Mitchell}
\affiliation{North Carolina State University, Raleigh, NC 27695, USA}
\affiliation{Triangle Universities Nuclear Laboratory, Durham, NC 27708, USA}
\author{J. Rekstad}
\affiliation{Department of Physics, University of Oslo, N-0316 Oslo, Norway}
\author{A. Schiller}
\affiliation{NSCL, Michigan State University, East Lansing, MI 48824, USA}
\author{S. Siem}
\affiliation{Department of Physics, University of Oslo, N-0316 Oslo, Norway}
\author {A. Voinov}
\affiliation{Department of Physics and Astronomy, Ohio University, Athens, OH 
45701, USA}

\date{\today}

\begin{abstract}
The $\cal{S}$ shape of the canonical heat capacity is known as a signature of 
the pairing transition and along an isotopic chain it is significantly more 
pronounced for nuclei with an even number of neutrons than with an odd number. 
Although the heat capacities extracted from experimental level densities in 
$^{93-98}$Mo exhibit a clear $\cal{S}$ shape, they do not show such an odd-even
staggering. To understand the underlying physics, we analyze thermal quantities
evaluated from the partition function calculated using the static-path plus 
random-phase approximation (SPA+RPA) in a monopole pairing model with 
number-parity projection. The calculated level densities reproduce very well 
the experimental data and they also agree with estimates using the back-shifted
Fermi-gas model. We clarify the reason why the heat capacities for Mo isotopes 
do not show odd-even staggering of the $\cal{S}$ shape. We also discuss thermal
odd-even mass differences in $^{94-97}$Mo using the three-, four-, and 
five-point formula. These thermal mass differences are regarded as indicators 
of pairing correlations at finite temperature. 
\end{abstract}

\pacs{21.60.Jz, 21.10.Ma, 05.30.-d}

\maketitle

Level density is one of the basic ingredients required for theoretical studies 
of nuclear structure and reactions. Recently, level densities for $^{93-98}$Mo 
\cite{Chankova} have been extracted by the group at the Oslo Cyclotron 
Laboratory. In their paper, the $\cal{S}$ shape of the canonical heat capacity 
is interpreted as consistent with a pairing phase transition with a critical 
temperature for the quenching of pairing correlations of 
$T_c\sim 0.7$--1.0~MeV\@. 

Pairing correlations are of special importance in nuclear physics. The 
Bardeen-Cooper-Schrieffer (BCS) theory \cite{BCS} has successfully described 
the sharp phase transition connected to the breakdown of pairing correlations 
for infinite Fermi system of electrons in low-temperature superconductors. This
sharp phase transition leads to a discontinuity of the heat capacity at the 
critical temperature, which indicates a second-order phase transition. For 
finite Fermi systems such as the atomic nucleus, however, nucleon number 
fluctuations and thermal and quantal fluctuations beyond the mean field become 
large. These fluctuations wash out the discontinuity of the heat capacity in 
the mean-field approximation and give rise to an $\cal{S}$ shape 
\cite{Rossignoli}\@. In recent theoretical approaches such fluctuations have 
been taken into account. Examples are: the static-path plus random-phase 
approximation (SPA+RPA) \cite{Rossignoli} and shell-model Monte-Carlo (SMMC) 
calculations \cite{Rombouts,Liu,Alhassid}\@. 

\begin{figure}[t!]
\includegraphics[width=8cm,height=10cm]{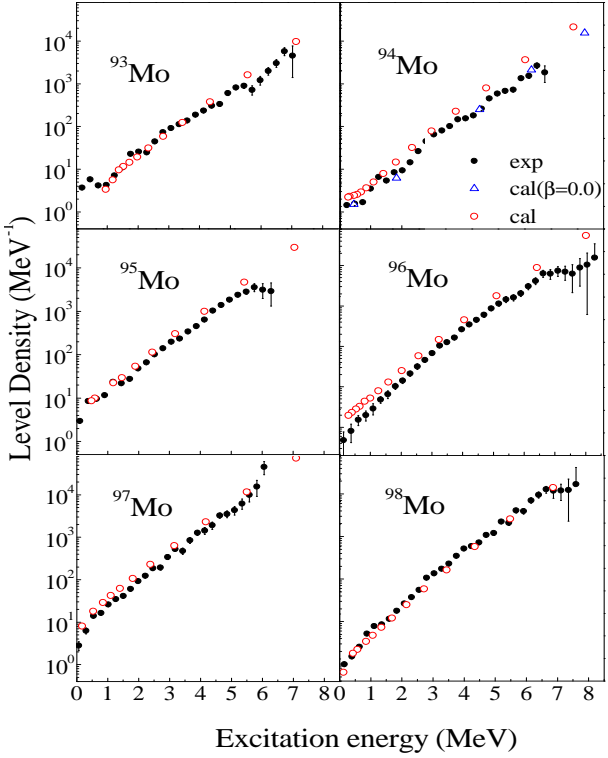}
\caption{(Color online) Experimental and calculated level densities as a 
function of excitation energy in $^{93-98}$Mo \protect\cite{Chankova}\@. The 
open circles denote the SPA+RPA calculations. The error bars show the 
statistical uncertainties.}
\label{fig1}
\end{figure}

\begin{figure}[t!]
\includegraphics[width=8cm,height=10cm]{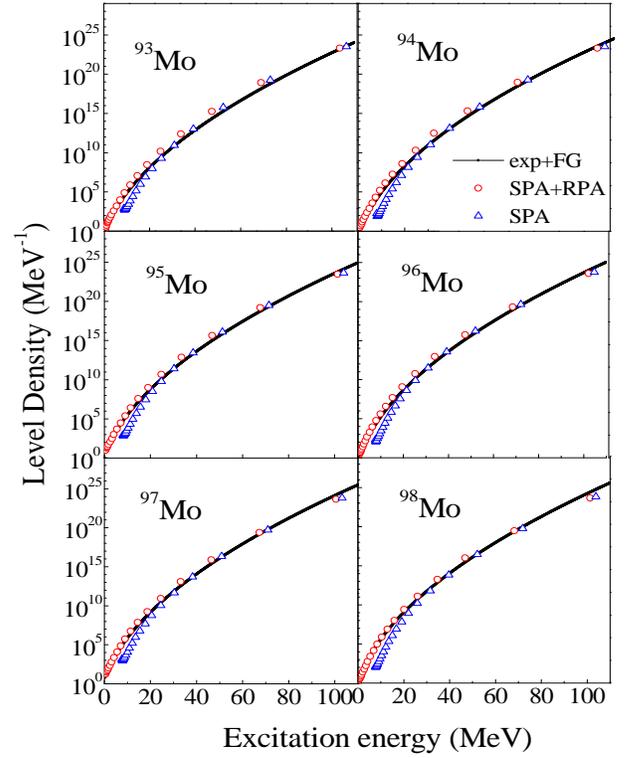}
\caption{(Color online) Extrapolated and calculated level densities as a 
function of excitation energy in $^{93-98}$Mo. The open circles and triangles 
denote the SPA+RPA and SPA calculations, respectively.}
\label{fig2}
\end{figure}

\begin{figure}[t!]
\includegraphics[width=8cm,height=10cm]{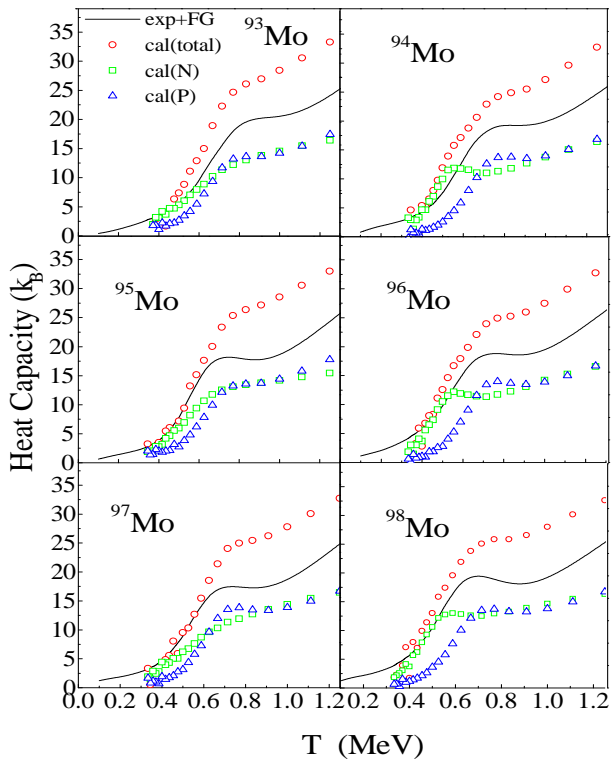}
\caption{(Color online) Heat capacities as a function of temperature $T$ in 
$^{93-98}$Mo. The solid curve denotes the heat capacity extracted from the 
experimental level density. The open squares and open triangles are the 
calculated heat capacities for neutrons and protons, respectively. The total 
heat capacities are represented by the open circles.}
\label{fig3}
\end{figure}

It has recently been reported \cite{Schiller,Melby} that the canonical heat 
capacities extracted from experimental level densities exhibit $\cal{S}$ shapes
with a peak around the critical temperature. The $\cal{S}$ shape is well 
correlated with the suppression of the number of spin-zero pairs around the 
critical temperature \cite{Liu,Alhassid}\@. The important feature is an 
odd-even staggering of the $\cal{S}$ shape where the $\cal{S}$ shape of the 
heat capacity for an even number of neutrons is significantly more pronounced 
than the one for an odd number of neutrons \cite{Liu,Rossignoli,Schiller}\@. A 
shell-model analysis \cite{Liu,Alhassid,Kaneko1} suggests that the difference 
between heat capacities of nuclei with even and odd numbers of neutrons is an 
important measure for understanding the breaking of neutron pairs. Figure 
\ref{fig1} shows the experimental level densities for $^{93-98}$Mo 
\cite{Chankova}\@. The extrapolated level densities using the back-shifted 
Fermi-gas model shown in Fig.\ \ref{fig2} make it possible to systematically 
study pairing correlations at finite temperatures in a series of odd- and 
even-mass nuclei. The heat capacities extracted from experimental level 
densities all show a clear $\cal{S}$ shape as seen in Fig.\ \ref{fig3}\@. 
Therefore, the heat capacities of $^{93-98}$Mo reveal a systematic, different 
from that of other experimental data \cite{Schiller,Melby} and theoretical 
calculations \cite{Liu,Alhassid,Rossignoli}\@.

The $\cal{S}$ shape of the heat capacity around the critical temperature is 
not a quantity which reflects details of the pairing transition. We cannot 
distinguish between, e.g., quenching of proton-proton ($pp$) and 
neutron-neutron ($nn$) pairing correlations from the $\cal{S}$ shape. In our 
recent papers \cite{Kaneko1,Kaneko2,Kaneko3}, we demonstrated that the 
suppression of pairing correlations around the critical temperature appears in 
thermal odd-even mass differences using the three-point formula. For heated 
systems, the thermal odd-even mass difference is the natural extension of the 
odd-even mass difference \cite{OES,Dobaczewski,Duguet} observed for nuclear 
ground-state masses.

The aim of this rapid communication is to investigate pairing properties in 
$^{93-98}$Mo. We report on SPA+RPA calculations in the monopole pairing model 
using a deformed Woods-Saxon potential with spin-orbit interaction \cite{Cwiok}
and we discuss physical reasons for why the heat capacities for Mo isotopes do 
not show an odd-even staggering of the $\cal{S}$ shape. 

Let us start from a monopole pairing Hamiltonian 
\begin{equation}
H=\sum_{k,\tau}\varepsilon_{k,\tau}(c_{k,\tau}^{\dag}c_{k,\tau}+c_{\bar{k},
\tau}^{\dag}c_{\bar{k},\tau})-\sum_{\tau}G_{\tau}P_\tau^{\dag}P_\tau, 
\label{eq:1}
\end{equation}
where $\tau=n,p$ and $\bar{k}$ denotes the time reversed state. Here, 
$\varepsilon_{k,\tau}$ is a single-particle energy and $P_\tau$ is the pairing 
operator $P_\tau=\sum_kc_{\bar{k},\tau}c_{k,\tau}$\@. Here, we neglect 
proton-neutron pairing interactions because in $^{93-98}$Mo the respective 
single-particle orbitals might be too far from each other to play a role. By 
means of the SPA+RPA \cite{Rossignoli} based on the Hubbard-Stratonovich 
transformation \cite{Hubbard}, the number-parity projected partition function 
is given by 
\begin{eqnarray}
Z^{\rm c}&=&{\rm Tr}\left[P_NP_Ze^{-H/T}\right]_{\rm SPA+RPA}\nonumber\\
&=&\prod_\tau\frac{2}{G_{\tau}T}\int_0^\infty\Delta_{\tau}d\Delta_{\tau}e^
{-\Delta_\tau^2/G_{\tau}T}Z_{\tau}C_{\rm RPA}^\tau 
\label{eq:2}
\end{eqnarray}
with
\begin{eqnarray}
Z_\tau&=&\frac{1}{2}\prod_ke^{-\gamma_{k,\tau}/T}\left(1+
e^{-\lambda_{k,\tau}/T}\right)^2\nonumber\\
&&\times\left[1+\sigma\prod_{k^\prime}\tanh^2(\lambda_{k',\tau}/T)\right]
\end{eqnarray}
and
\begin{equation}
C_{\rm RPA}^\tau=\prod_k\frac{\omega_{k,\tau}\sinh[\lambda_{k,\tau}/T]}
{2\lambda_{k,\tau}\sinh[\omega_{k,\tau}/2T]}
\label{eq:3}
\end{equation}
where, instead of the exact number projection, we introduced the number-parity 
projections $P_N=(1+\sigma e^{i\pi N})/2$ for neutron number $N$ and 
$P_Z=(1+\sigma e^{i\pi Z})/2$ for proton number $Z$ ($\sigma$ denotes the even 
or odd number parity \cite{Rossignoli})\@. It has recently been shown that in 
the monopole pairing case the SPA+RPA with number-parity projection reproduces 
well exact results \cite{Rossignoli}\@. The number-parity projection is 
essential to describe thermal properties at low temperature. Here, we use the 
notation $\omega_{k,\tau}$ for the conventional thermal RPA energies and 
$\lambda_{k,\tau}=\sqrt{\varepsilon^{\prime 2}_{k,\tau}+\Delta_\tau^2}$, 
$\varepsilon^\prime_{k,\tau}=\varepsilon_{k,\tau}-\mu_\tau-G_\tau/2$, and 
$\gamma_{k,\tau}=\varepsilon_{k,\tau}-\mu_\tau-\lambda_{k,\tau}$\@. The SPA 
partition function is obtained by neglecting the RPA partition function 
$C_{\rm RPA}^\tau$\@. The thermal energy can be calculated from 
$E=-\partial\ln Z^{\rm c}/\partial\beta$ where $\beta=1/T$\@. In this work, we 
use the single-particle energies $\varepsilon_{k,\tau}$ given by an axially 
deformed Woods-Saxon potential with spin-orbit interaction. We choose the 
Woods-Saxon parameters of \cite{Cwiok}, where $V_0=51.0$~MeV, $a=0.67$~fm, 
$\kappa=0.67$, $\lambda=20.3$, and $r_0=1.27$~fm. We have calculated the 
single-particle spectrum of this potential, and used it to compute the 
number-parity projected partition function. The deformation parameters for the 
even-even nuclei $^{94,96,98}$Mo are estimated from the experimental 
$B(E2;2_1^+\rightarrow 0_1^+)$ values, and are determined as $\beta_2=0.15$, 
0.17, and 0.17, respectively. Those for the odd nuclei $^{93,95,97}$Mo are 
chosen as $\beta_2=0.10$, 0.08, and 0.17, respectively. The 25 and 30 doubly 
degenerate single-particle levels with negative energy are taken with respect 
to neutrons and protons outside the $^{48}$Ca core. The positive energy levels 
(resonances and continuum states) are neglected \cite{Alhassid}\@. We adjusted 
the pairing force strengths at $G_n=22/A$~MeV for neutrons and $G_p=26.5/A$~MeV
for protons in order to reproduce the three-point odd-even mass differences. 

The level density can be evaluated from an inverse Laplace transformation of 
the partition function $Z^{\rm c}$ in the saddle-point approximation as  
\begin{equation}
\rho(E)\approx\frac{Z^{\rm c}e^{\beta E}}{[2\pi\partial^2\ln Z^{\rm c}/\partial
\beta^2]^{1/2}}. 
\label{eq:5}
\end{equation}
The calculated level densities as a function of thermal energy are shown in 
Fig.\ \ref{fig1}, where the calculated level density is also plotted for 
$\beta_2=0$\@. The calculations reproduce very well the experimental data. It 
should be noted that, more precisely, Eq.\ (\ref{eq:5}) gives the state density
and not the level density because the partition function $Z^{\rm c}$ in Eq.\ 
(\ref{eq:2}) includes $m$-degeneracy \cite{Chankova}\@. This difference may 
have an impact on the canonical heat capacity. 

\begin{center}
\begin{table}[t!]
\caption{Parameters used for the back-shifted Fermi-gas level density.}
\begin{tabular*}{85mm}{@{\extracolsep{\fill}}l|ccc|c}\hline\hline 
Nucleus  &$E_{\rm pair}$&$a$         &$C_1$ &$\eta$\\
         &(MeV)         &(MeV$^{-1}$)&(MeV) &      \\\hline
         &              &            &      &      \\
$^{98}$Mo&2.080         &11.33       &-1.521&0.87  \\
$^{97}$Mo&0.995         &11.23       &-1.526&0.65  \\
$^{96}$Mo&2.138         &11.13       &-1.531&0.46  \\
$^{95}$Mo&1.047         &11.03       &-1.537&0.34  \\
$^{94}$Mo&2.027         &10.93       &-1.542&0.25  \\
$^{93}$Mo&0.899         &10.83       &-1.547&0.08  \\\hline\hline
\end{tabular*}
\label{table1}
\end{table}
\end{center}

We now extrapolate the experimental level densities by the back-shifted 
Fermi-gas model of \cite{Gilbert,Egidy} 
\begin{equation}
\rho_{\rm BSFG}(E)=\eta\frac{\exp\left[2\sqrt{aU}\right]}{12\sqrt{2}a^{1/4}
U^{5/4}\sigma_I}, 
\label{eq:6}
\end{equation}
where the back-shifted energy is $U=E-E_1$ and the spin-cutoff parameter 
$\sigma_I$ is taken as $\sigma_I^2=0.0888A^{2/3}\sqrt{aU}$\@. The level-density
parameter $a$ and the parameter $E_1$ are given by $a=0.21A^{0.87}$~MeV$^{-1}$ 
and $E_1=C_1+E_{\rm pair}$, respectively, where the back-shift is parameterized
by $C_1=-6.6A^{-0.32}$~MeV and the pairing energy $E_{\rm pair}$ is based on 
pairing-gap parameters $\Delta_n^{(3)}$ and $\Delta_p^{(3)}$ evaluated from 
odd-even mass differences \cite{Dobaczewski}\@. The factor $\eta$ is introduced
in order to reproduce experimental neutron-resonance spacings. The parameters 
used for $^{93-98}$Mo are listed in Table \ref{table1}\@. Figure \ref{fig2} 
displays the extrapolated and calculated level densities as a function of 
excitation energy. The level densities from the SPA+RPA calculation are in good
agreement with the extrapolated ones. In Fig.\ \ref{fig2}, we also compare the 
SPA+RPA with the pure SPA calculation. It is noted, that because of the RPA 
contributions, the calculated level density is significantly increased below an
excitation energy of $\sim 20$~MeV\@.

\begin{figure}[t!]
\includegraphics[width=8cm,height=8cm]{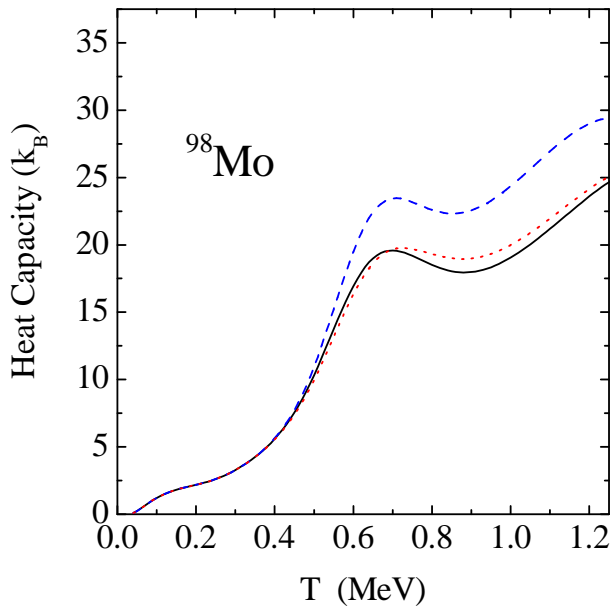}
\caption{(Color online) Heat capacities derived using different extrapolations 
of the experimental level-density curve in $^{98}$Mo. The extrapolations are 
performed with a 20\% increase of (i) the level-density parameter (dashed line)
or (ii) the pairing-gap parameter (dotted line), respectively. The solid curve 
indicates the heat-capacity curve of Fig.\ \protect\ref{fig3}\@.}
\label{fig4}
\end{figure}

To study thermal properties derived from experimental level densities, let us 
start from the partition function in the canonical ensemble, i.e., the Laplace 
transform of the level density $\rho(E_i)$ 
\begin{equation}
Z(T)=\sum_{i=0}^\infty\delta E_i\rho(E_i){\rm e}^{-E_i/T}, 
\label{eq:7}
\end{equation}
where $E_i$ are the excitation energies and $\delta E_i$ are the energy bins. 
Then, the thermal energy is expressed as
\begin{equation}
E(Z,N,T)=\sum_{i=0}^\infty\delta E_i\rho(E_i)E_i{\rm e}^{-E_i/T}/Z(T), 
\label{eq:8}
\end{equation}
and the heat capacity is given by 
\begin{equation}
C(Z,N,T)=\frac{\partial E(Z,N,T)}{\partial T}. 
\label{eq:9}
\end{equation}
Formally, the calculations using Eqs.\ (\ref{eq:7})--(\ref{eq:9}) require an 
infinite summation. However, the experimental level densities in Fig.\ 
\ref{fig1} only cover the excitation energy up to 6.0--8.5~MeV\@. Since the 
calculated level densities agree well with those extrapolated using the 
back-shifted Fermi-gas model of Eq.\ (\ref{eq:6}), we use this extrapolation to
extend the experimental level-density data up to an excitation energy of 
$\sim$180~MeV before evaluating Eqs.\ (\ref{eq:7})--(\ref{eq:9})\@. The 
$\cal{S}$-shape behavior of the heat capacities for $^{93-98}$Mo can be 
reproduced by the calculation, although the calculated heat capacities are 
larger than the experimental ones above the critical temperature. The 
characteristic feature is that all of the nuclei $^{93-98}$Mo exhibit a clear 
$\cal{S}$ shape of the heat capacity around the critical temperature 
$T_c\sim 0.7$--0.9~MeV\@. The calculated heat capacities are divided into 
neutron and proton parts as shown in Fig.\ \ref{fig3}\@. This figure also 
indicates that the heat capacity for the neutron part shows a different 
behavior from the one for the proton part. For the neutron part, the critical 
temperature is $\sim 0.55$~MeV and the $\cal{S}$ shape of the heat capacity for
an even number of neutrons is more pronounced than the one for an odd number of
neutrons. For the proton part, on the other hand, the heat capacities show the 
same clear $\cal{S}$ shape around $T_c\sim 0.75$~MeV for all of the nuclei 
$^{93-98}$Mo. Thus, we conclude that the experimentally observed $\cal{S}$ 
shape of the heat capacity in $^{93-98}$Mo can be attributed mainly to the 
proton contribution. 

We will now discuss how sensitive the extracted heat capacities in Fig.\ 
\ref{fig3} are with respect to the level-density extrapolation. Figure 
\ref{fig4} shows the extracted heat capacities using a 20\% larger (i) 
level-density parameter $a$ or (ii) pairing-gap parameter $E_{\rm pair}$\@. The
increased $a$ makes the slope of the heat-capacity curve steeper than before; 
the increase of the pairing-gap parameter does not change the heat-capacity 
curve significantly. In either case, the heat capacity shows a pronounced 
$\cal{S}$ shape which indicates that this qualitative feature is very robust 
with respect to changes in the Fermi-gas parameters of the level-density 
extrapolation.

\begin{figure}[t!]
\includegraphics[width=8cm,height=10cm]{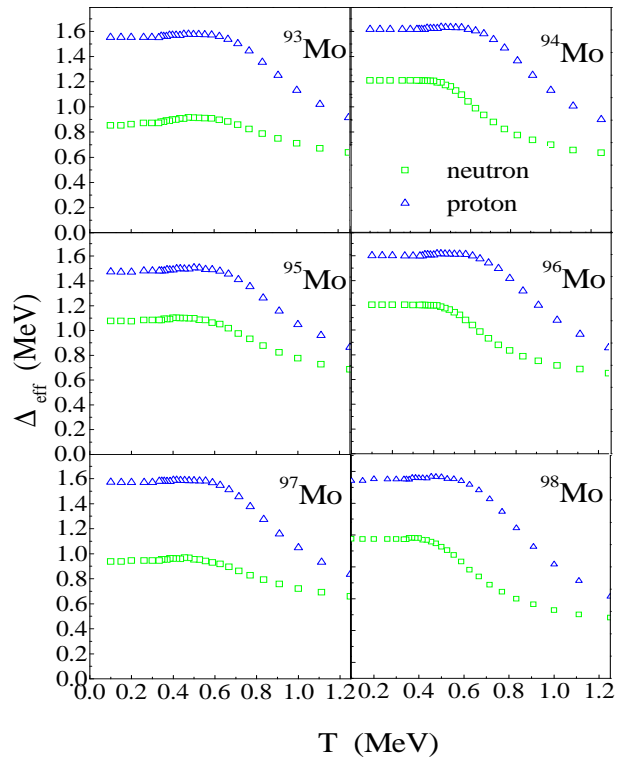}
\caption{(Color online) Effective pairing gap as a function of temperature. The
open squares and open triangles denote the effective pairing gaps for neutrons 
and protons, respectively.}
\label{fig5}
\end{figure}

\begin{figure}[t!]
\includegraphics[width=8cm,height=10cm]{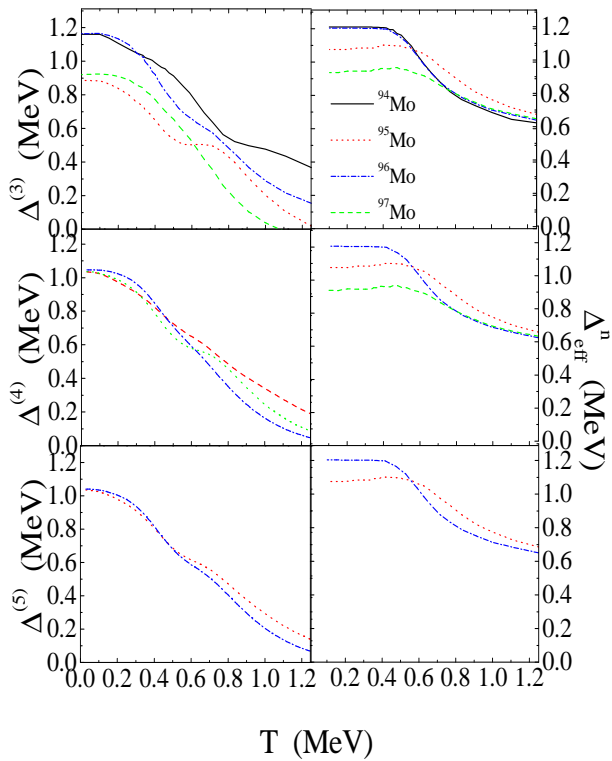}
\caption{(Color online) Left: thermal pairing gaps extracted from the three-, 
four-, and five-point indicators of the thermal odd-even mass difference as a 
function of temperature. Right: effective pairing gaps for neutrons in the 
monopole-pairing-model calculations of this work.}
\label{fig6}
\end{figure}

The $\cal{S}$ shape has been discussed to be correlated with the breaking of 
nucleon Cooper pairs \cite{Liu,Kaneko1}\@. Therefore, we further investigate 
the pairing properties in the calculations. Figure \ref{fig5} shows the 
effective pairing gap defined by 
\begin{equation}
\Delta^\tau_{\rm eff}=G_\tau\left[\frac{1}{\beta}\frac{\partial\ln Z^{\rm c}}
{\partial G_\tau}\right]^{1/2}.
\label{eq:10}
\end{equation}
Around $T_c\sim 0.6$~MeV, $\Delta^n_{\rm eff}$ decreases more rapidly for an 
even number of neutrons than for an odd number of neutrons. The suppression of 
$\Delta^n_{\rm eff}$ is well correlated with the $\cal{S}$ shape of the heat 
capacity for neutrons in Fig.\ \ref{fig3}\@. Furthermore, $\Delta^p_{\rm eff}$ 
decreases drastically around $T_c\sim 0.8$~MeV for all of the nuclei 
$^{93-98}$Mo. This suppression of $\Delta^p_{\rm eff}$ correlates well in 
temperature with the pronounced $\cal{S}$ shape of the proton heat capacity in 
Fig.\ \ref{fig3}\@. Thus, the $\cal{S}$ shape in Fig.\ \ref{fig3} can be 
understood in terms of the suppression of the effective pairing gap. 

The ground-state odd-even mass difference is known to be a measure of pairing 
correlations \cite{OES,Duguet}\@. Extending this odd-even mass difference, we 
have proposed the thermal odd-even mass difference as a measure of pairing 
correlations at finite temperatures \cite{Kaneko1,Kaneko2}\@. The three-point 
thermal odd-even mass difference for neutrons is given by 
\begin{eqnarray}
\lefteqn{\Delta_n^{(3)}(Z,N,T)=\frac{(-1)^N}{2}\left[B_t(Z,N+1,T)\right.}
\nonumber\\
&&\left.-2B_t(Z,N,T)+B_t(Z,N-1,T)\right],
\label{eq:11}
\end{eqnarray}
where the thermal energy $B_t$ is defined by $B_t(Z,N,T)=E(Z,N,T)+B(Z,N)$ and 
$B(N,Z)$ is the binding energy at zero temperature. The thermal odd-even mass 
difference $\Delta_n^{(3)}$ is evaluated using $E(Z,N,T)$ from Eq.\ 
(\ref{eq:8})\@. In our previous work \cite{Kaneko2}, we extracted the thermal 
odd-even mass difference $\Delta_n^{(3)}$ in $^{184}$W from the experimental 
level densities of $^{183}$W, $^{184}$W, and $^{185}$W, and we obtained a 
drastic decrease of $\Delta_n^{(3)}$ at the critical temperature, a signal 
which can be correlated to a corresponding $\cal{S}$ shape of the heat 
capacity. As we have already mentioned in Ref.\ \cite{Kaneko2}, the sudden 
decrease of the thermal odd-even mass differences is interpreted as a rapid 
breaking of nucleon Cooper pairs. It is therefore interesting to see whether 
such a drastic change is found for the Mo isotopes as well. 

The upper left panel of Fig.\ \ref{fig6} shows the thermal odd-even mass 
differences $\Delta_n^{(3)}$ for neutrons as a function of temperature. For 
$^{94}$Mo$, \Delta_n^{(3)}$ starts to decrease rather suddenly around 
$T=$0.6~MeV, however, this decrease is never as drastic as the one observed for
$^{184}$W \cite{Kaneko2}\@. Moreover, for $^{95-97}$Mo the thermal odd-even 
mass differences $\Delta_n^{(3)}$ do not display any sudden onset of quenching,
rather they exhibit a more monotonous decrease with increasing temperature 
starting as early as around $T=$0.2~MeV\@. We now compare these thermal 
odd-even mass differences $\Delta_n^{(3)}$ to the effective pairing gaps 
$\Delta^n_{\rm eff}$ for neutrons in the monopole-pairing-model calculations 
for $^{94-97}$Mo (upper right panel of Fig.\ \ref{fig6})\@. The calculated 
effective pairing gaps decrease suddenly around the critical temperature 
corresponding to the presence of the $\cal{S}$ shape in the heat-capacity 
curves. At high temperatures of $T=$1.25~MeV, their values are 
$\Delta^n_{\rm eff}=$0.6--0.7~MeV which is in reasonable agreement with the 
thermal odd-even mass differences for $^{94}$Mo, however, the thermal odd-even 
mass differences for $^{95-97}$Mo are much smaller than the theoretical 
estimates at these high temperatures.

It has been demonstrated recently that $\Delta_n^{(3)}$ contains additional 
mean-field contributions when realistic pairing forces are used in the 
calculations \cite{Duguet}, however, the five-point odd-even mass difference 
$\Delta_n^{(5)}$ extracts the pairing gap. Therefore, we also calculate the 
four- and five-point thermal odd-even mass differences 
\begin{equation}
\Delta^{(4)}_n(Z,N,T)=\frac{1}{2}\left[\Delta^{(3)}_n(Z,N,T)+
\Delta^{(3)}_n(Z,N-1,T)\right],
\label{eq:12}
\end{equation}
and
\begin{eqnarray}
\lefteqn{\Delta^{(5)}_n(Z,N,T)=\frac{1}{4}\left[\Delta^{(3)}_n(Z,N+1,T)\right.}
\nonumber\\
&&+\left.2\Delta^{(3)}_n(Z,N,T)+\Delta^{(3)}_n(Z,N-1,T)\right]. 
\label{eq:13}
\end{eqnarray}
The lower left panels of Fig.\ \ref{fig6} show the thermal odd-even mass 
differences for neutrons $\Delta_n^{(4,5)}$ as a function of temperature, where
$\Delta_n^{(4)}$ and $\Delta_n^{(5)}$ are obtained by the same treatment as 
$\Delta_n^{(3)}$\@. The thermal odd-even mass differences $\Delta_n^{(4,5)}$ 
extracted from the experimental data also show a monotonous decrease in 
contrast to the sudden quenching of the effective pairing gap around the 
critical temperature in the monopole-pairing-model calculations for 
$^{94-97}$Mo (lower right panels)\@. 

Thus, the majority of the thermal odd-even mass differences 
$\Delta_n^{(3,4,5)}$ decrease monotonously and do not show any sudden 
quenching, while the theoretical calculations suggest the presence of a pairing
phase transition at a critical temperature of $T=$0.6~MeV\@. To possibly 
explain this discrepancy, we would like to point out that the thermal odd-even 
mass differences $\Delta_n^{(3,4,5)}$ are phenomenological indicators extracted
from level densities; they may include other correlations than just monopole 
pairing correlations. E.g., quadrupole correlations, which are not taken into 
account in our calculations, could be important for understanding thermal 
properties. Such additional seniority non-conserving correlations may wash out 
the expected sudden decrease of $\Delta_n^{(3,4,5)}$ due to the breaking of 
nucleon Cooper pairs at the critical temperature. Hence, the relation between 
the odd-even mass differences and the pairing correlations at finite 
temperature still remains an open question. 

In conclusion, we have investigated pairing properties in the $^{93-98}$Mo 
isotopic chain of neutron-odd and -even nuclei by performing SPA+RPA 
calculations within a monopole pairing model. The calculations reproduce very 
well the experimental level densities, and explain an unusual feature recently 
found for $^{93-98}$Mo, namely that the corresponding heat capacities do not 
show any pronounced odd-even staggering. The thermal three-, four-, and 
five-point odd-even mass differences in $^{94-97}$Mo, which are regarded as 
measures of pairing correlations at finite temperature, were extracted from the
experimental level densities. They show a monotonous decrease with increasing 
temperature and no sudden drastic quenching. Further theoretically studies in 
this direction are in progress. On the experimental side, we plan for further 
experiments to also extract proton odd-even mass differences at finite 
temperature. Strong quenching of proton pairing correlations as seen on Fig.\ 
\ref{fig5} suggests a drastic suppression of odd-even mass differences for 
isotonic chains.

\end{document}